\documentstyle[sprocl,epsfig]{article}

\def\Journal#1#2#3#4{{#1} {\bf #2}, #3 (#4)}

\def\NPB{{\em Nucl. Phys.} B}

\def\PRL{\em Phys. Rev. Lett.}
\def\PRD{{\em Phys. Rev.} D}

\def\be{\begin{equation}}
\def\ee{\end{equation}}
\def\bea{\begin{eqnarray}}
\def\eea{\end{eqnarray}}

\begin{document}

\title{EVIDENCE FOR DUAL SUPERCONDUCTIVITY OF QCD GROUND STATE.}

\author{A. DI GIACOMO}

\address{Dip. Fisica Universit\`a and INFN, Via Buonarroti 2 ed.B 56126
PISA, ITALY\\E-mail: digiacomo@pi.infn.it}


\maketitle\abstracts{A discussion is made of the strategy to check dual
superconductivity of the vacuum as a mechanism of colour confinement. Recent
evidence from lattice is reviewed.}
\section{Introduction}
No reliable analytic approach exists to QCD at large distances. The usual
perturbative quantization leads to an $S$ matrix which is not Borel summable.
For reasons which are not understood the perturbative expansion works anyhow at
small distances, where a few terms correctly describe experiments. It fails at
large distances, where the coupling is large, in particular in describing
confinement of colour.

Attempts have been made to describe the degrees of freedom relevant to
confinement by effective models. Particularly attractive from the theoretical
point of view, is the possibility that vacuum behaves as a dual
superconductor\cite{1,2}.
Dual Meissner effect would accordingly produce confinement by constraining the
chromoelectric field into Abrikosov flux tubes, with energy proportional to
their length.

The mechanism is appealing because it relies on a symmetry property.
Superconductivity is a Higgs mechanism, by which a charged field acquires a non
zero v.e.v., the order parameter in the Landau Ginzburg free energy. The ground
state has no definite charge, the $U(1)$ related to charge conservation
being spontaneously broken.

For QCD magnetic charges should condense in the confined phase, and break some
magnetic $U(1)$ symmetry. A dual order parameter, a disorder parameter in the
language of statistical mechanics, would then describe this change of symmetry.

Only a non perturbative quantization, like lattice, can help in cheking if the
above mechanism is at work. The simplest strategy to do that consists of two
steps
\begin{itemize}
\item[1.] Identify the relevant magnetic $U(1)$.
\item[2.] Check by a disorder parameter if it breaks spontaneously.
\end{itemize}
\section{Identifying monopoles: the abelian projection.}
Monopoles in non abelian gauge theories were first discovered\cite{3,4} as
solitons in the Higgs phase in a gauge theory with gauge group $SO(3)$ coupled
to a scalar field in the adjoint representation\cite{5}
\begin{equation}
{\cal L} = -\frac{1}{4}\vec G_{\mu\nu}\vec G_{\mu\nu} +
(D_\mu\vec\Phi)^*(D_\mu\vec\Phi) -
\frac{\lambda}{2}(\vec\Phi^2-\frac{\mu^2}{\lambda})^2 \label{eq:1}\end{equation}
It was shown in ref's\cite{3,4} that monopoles exist as static solutions
(solitons) in the Higgs phase of the model, i.e. for $\mu^2 > 0$,
$\vec\Phi_0 = \langle\vec\Phi\rangle\neq 0$.

In the hedgehog gauge the monopole has the form
\begin{equation}
\vec\Phi(\vec r) = f(r)\Phi_0\hat r\quad \vec A_0(\vec r) = 0\quad
(A_i)^a = \frac{2}{g}\varepsilon_{iak} \frac{\hat r^k}{r} h(r)
\label{eq:2}\end{equation}
with $f(r), h(r) \sim 1$ as $r\gg 1/\mu$.

A gauge transformation to the unitary gauge, $U(\vec r)$,
\begin{equation}
U(\vec r) \hat\Phi(\vec r) = \hat \Phi_0 \equiv (0,0,1)
\label{eq:3}\end{equation}
is defined up to a residual $U(1)$ gauge group of rotations around the $z$
axis. $U(\vec r)$ is singular at the zero of $\vec \Phi(\vec r)$, $\vec r = 0$.
 $U(\vec r)$ is usually called an abelian projection. For the
monopole solution the abelian field of
the residual $U(1)$ in the abelian projected gauge
\begin{equation}
{\cal F}_{\mu\nu} = \partial_\mu A^3_\nu - \partial_\nu A^3_\mu
\label{eq:4}\end{equation}
is the field of a Dirac monopole.

${\cal F}_{\mu\nu}$ can be written in a gauge invariant form as
\begin{equation}
{\cal F}_{\mu\nu} = \hat\Phi \vec G_{\mu\nu} -
\frac{1}{g}\hat\Phi(D_\mu\hat\Phi
\wedge D_\nu\hat \Phi)\label{eq:5}\end{equation}
Calling ${\cal F}_{\mu\nu}^* = \frac{1}{2} \varepsilon_{\mu\nu\rho\sigma}
{\cal F}^{\rho\sigma}$, $j^M_\nu = \partial^\mu {\cal F}^*_{\mu\nu}$
we have identically
\begin{equation}
\partial^\nu j^M_\nu = 0 \label{eq:6}\end{equation}
Eq.(\ref{eq:6}) identifies an $U(1)$ magnetic symmetry. The corresponding
charge $Q$ is a colour singlet
and is equal to two magnetic units for the monopole solution.
Also ${\cal F}_{\mu\nu}$ and ${\cal F}^*_{\mu\nu}$ are colour singlets.

More generally, an abelian projection $U(\vec r)$ can be performed which is
defined by eq.(\ref{eq:3}) on a generic configuration. $U(\vec r)$ is
singular at the zeros of $\vec \Phi(\vec r)$. Around these points the field has
the topology of an abelian monopole\cite{6}.

We could think of a slightly more general model in which an additional
Higgs field
$\vec\Phi'$ is present, e.g. with the same potential as $\vec\Phi$
\begin{eqnarray}
{\cal L} &=& -\frac{1}{4}\vec G_{\mu\nu}\vec G_{\mu\nu} +
(D_\mu\vec\Phi)^*(D_\mu\vec\Phi) -
\frac{\lambda}{2}(\vec\Phi^2-\frac{\mu^2}{\lambda})^2 \label{eq:7}\\
&& +
(D_\mu\vec\Phi')^*(D_\mu\vec\Phi') -
\frac{\lambda}{2}(\vec\Phi'^2-\frac{\mu^2}{\lambda})^2 \nonumber
\end{eqnarray}
We can define an abelian projection which brings $\vec\Phi$ to the unitary gauge,
as in the simple model above, and define the gauge invariant field
${\cal F}_{\mu\nu}^*$, and the
corresponding magnetic $U(1)$. We can play the same
game with $\vec\Phi'$, and this will in general bring to a different abelian
projection and to a different $U(1)$. Both
magnetic charges
are gauge invariant. On a
given field configuration the zeros of $\vec\Phi$ and $\vec\Phi'$ will not
coincide in general so that the two abelian projections define different
monopoles. However the theory is totally symmetric under the exchange
$\vec\Phi\leftrightarrow\vec\Phi'$ and hence the two monopole species defined
by the two abelian projections must be physically equivalent.

There is in the literature a misuse of terminology: the abelian projections
are named from the abelian projected gauge, so that the two monopole species
defined in the above example are called monopole in the gauge $\vec\Phi$ and
monopoles in the  gauge $\vec\Phi'$: a possible difference of physics, e.g. if
the two fields have a different potential
 is called gauge dependence. This
terminology is misleading: usually, e.g. in QED, as gauge dependent
is  meant a quantity
 which does not depend only on the physical fields, $F_{\mu\nu}$, but could
depend on the choice of the gauge. Monopole charges defined by any abelian
projection are instead physically well defined and gauge invariant quantities.

In QCD there is no Higgs field. However there exist infinitely many fields
transforming in the adjoint representation, and each of them
can define an
abelian projection and with it  a monopole species.

On the lattice any parallel transport along an arbitrary path $C$ coming back
to the starting point defines an abelian projection and a monopole species.
The corresponding monopoles are different in number and located in different
sites, configuration by configuration. A possible guess is that they are all
physically equivalent\cite{6}, in the same
way as the two monopole species of
the model eq.(\ref{eq:7}). For each of them it is anyhow possible to investigate
condensation in the vacuum and dual superconductivity.
Some results will be presented in the next section.

An alternative attitude is that some abelian projection is better than others.
This attitude is popular among the practitioners of the so called maximal abelian gauge.
This is an abelian projection for which the operator $\vec\Phi$ is implicitly
defined by maximizing numerically  the quantity
\begin{equation}
A = \sum_{x,n} Tr\left[
\Omega(x) U_\mu(x) \Omega^\dagger(x+\hat\mu) \sigma_3
\Omega(x+\mu) U^\dagger(x) \Omega^\dagger(x) \sigma_3\right]
\label{eq:8}\end{equation}
with respect to the gauge transformation $\Omega(x)$.

The numerical output is that in the new gauge  all the links $U_\mu(x)$ are
practically aligned along $\sigma_3$, within $10-20\%$. A remarkable
observation\cite{7} which is a consequence of this fact is the so called
``abelian dominance''. Quantities like e.g. the string tension, when computed
in the $U(1)$ residual gauge, agree within $10-20\%$ with the exact result. In
addition the abelian monopole part, corresponding to integer number of $2\pi$
in the abelian plaquettes, saturates the abelian approximation to within $90\%$
again.

Dominance is interpreted as special relevance of the specific monopoles in the long range
physics.

From theoretical point of view we find more significant and anyhow necessary
to investigate the
symmetry of the vacuum, i.e. the condensation of different monopole species in
connection with confinement.
\section{Detecting dual superconductivity.}
In the language of statistical mechanics the main issue of the problem is
duality: the gauge field of monopoles presents non trivial connection or topology.
A creation operator for monopoles has the form of a translation
of the field in the Schr\"odinger representation by a monopole configuration.
In $U(1)$ gauge theory\cite{8}
\[A_0^{mon}(\vec x, \vec y) = 0\qquad \vec A^{mon}(\vec x, \vec y) =
\frac{m}{g} \frac{\vec n\wedge (\vec x - \vec y)}{|\vec x - \vec y|^2
(|\vec x - \vec y| - (\vec x - \vec y)\cdot n)}\]
\begin{equation}
\mu(\vec y,t) =
\exp\left[ i \int d^3\vec x \vec E(\vec x,t) \vec A^{mon}(\vec x, \vec y)
\right]
\label{eq:9}\end{equation}
$\mu$ carries non zero magnetic charge.

Eq.(\ref{eq:9}) is the analog of the elementary translation
\begin{equation}
e^{i p a} |x\rangle = |x + a\rangle \label{eq:10}\end{equation}
$\vec E$ is the conjugate momentum to the field.
Some technical modifications will be needed to keep the compactness of the
theory into account\cite{8} on the lattice, and some extra care to perform the shift in
the abelian projected $U(1)$ for non abelian gauge theory\cite{9}.

$\langle\mu\rangle$ is then measured. $\langle\mu\rangle\neq 0$ means
spontaneous breaking of magnetic $U(1)$, and hence dual superconductivity.

For different monopole species we have measured $\langle\mu\rangle$ or better
$\rho = \frac{d}{d \beta}\ln\langle\mu\rangle$, as a function of temperature,
on asymmetric lattices $N_S \gg N_T$. $\rho$ contains the same information as
$\langle\mu\rangle$ and has less numerical problems in its determination.

Since $\langle\mu\rangle_{\beta=0} = 1$,
\begin{equation}
\langle\mu\rangle = \exp\left(\int_0^\beta \rho(x) d x\right)
\label{eq:10}\end{equation}
The typical behaviour of $\rho$ vs $\beta = 2 N_c/g^2$ is shown in fig.1 for
$SU(2)$ and different abelian projections, and for $SU(3)$ in fig.2, and fig.3\cite{9}.
\par\noindent
\begin{minipage}{0.95\linewidth}
\epsfxsize0.95\linewidth
{
\epsfbox{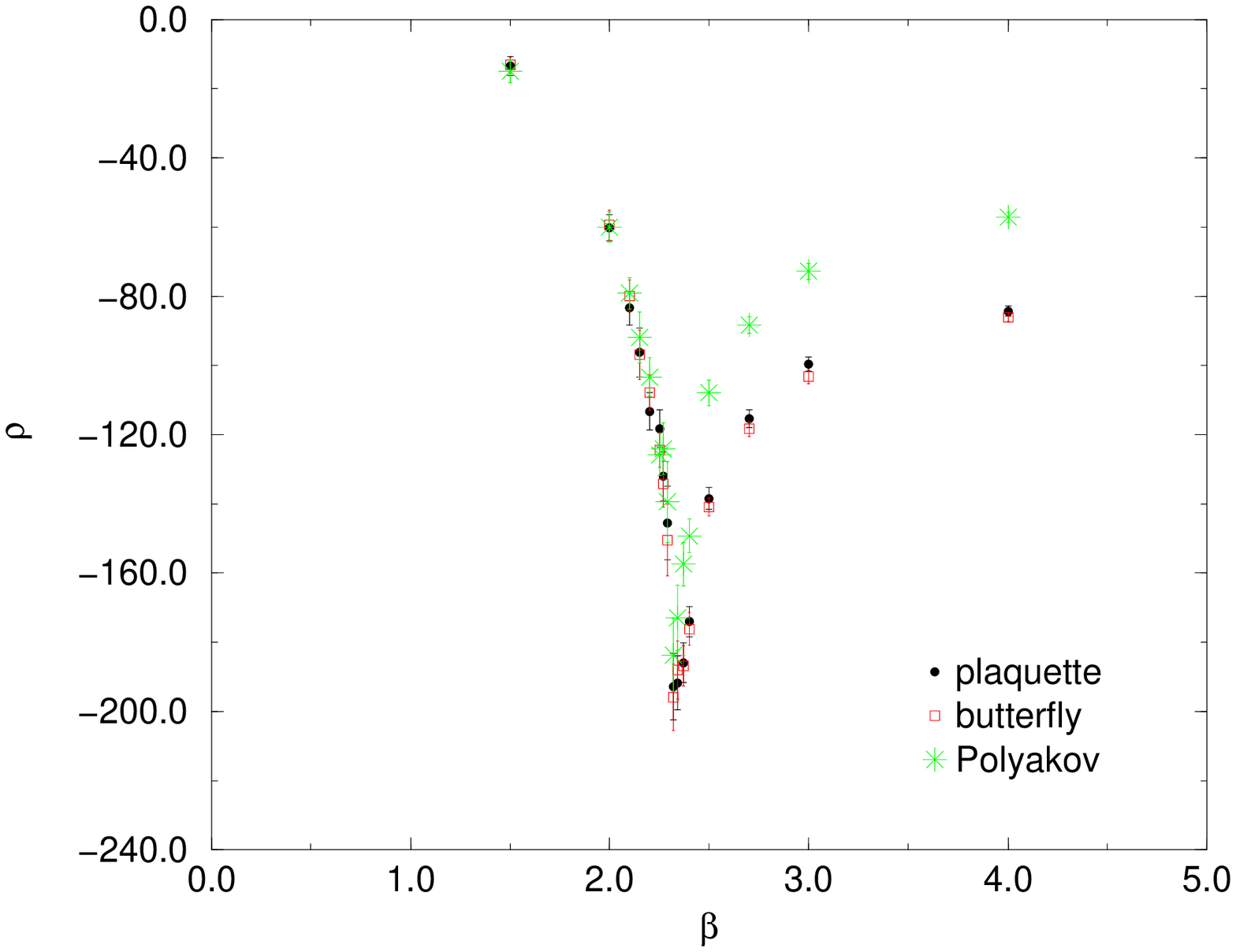}}
\end{minipage}\\
{\bf Fig.1} $\rho$ vs $\beta$ for different abelian projections
in $SU(2)$. The negative peak signals phase transition.
\vskip0.1in\par\noindent
There is no practical difference between different abelian projections, in
agreement with the guess of t'Hooft's\cite{6}.

The strong negative peak occurs at the deconfining transition, and, by
eq.(\ref{eq:10}), indicates a rapid drop to zero of $\langle\mu\rangle$.

At large $\beta$'s $\rho$ is computed by perturbation theory giving at the
leading order $\rho = - c_1 L_S + c_2$. As the spatial size $L_S\to\infty$,
$\rho\to -\infty$ and $\langle\mu\rangle = 0$, as expected for any disorder
parameter in the thermodynamical limit. For $\beta \sim \beta_c$ a finite size
scaling analysis with respect to $L_S$ can be performed.
Since the transition is second order for $SU(2)$ and weak first order
for $SU(3)$, the
correlation length $\xi$ goes large at
$\beta_c$, with some effective critical index $\nu$
\begin{equation}
\xi \sim(\beta_c-\beta)^{-\nu} \label{eq:11}\end{equation}
By dimensional arguments
\begin{equation}
\mu = \mu(\frac{a}{\xi},\frac{\xi}{L})\mathop\simeq_{\beta\to\beta_c}
\mu(0,\frac{\xi}{L})\label{eq:12}\end{equation}
\par\noindent
\begin{minipage}{0.95\linewidth}
\epsfxsize0.95\linewidth
{
\epsfbox{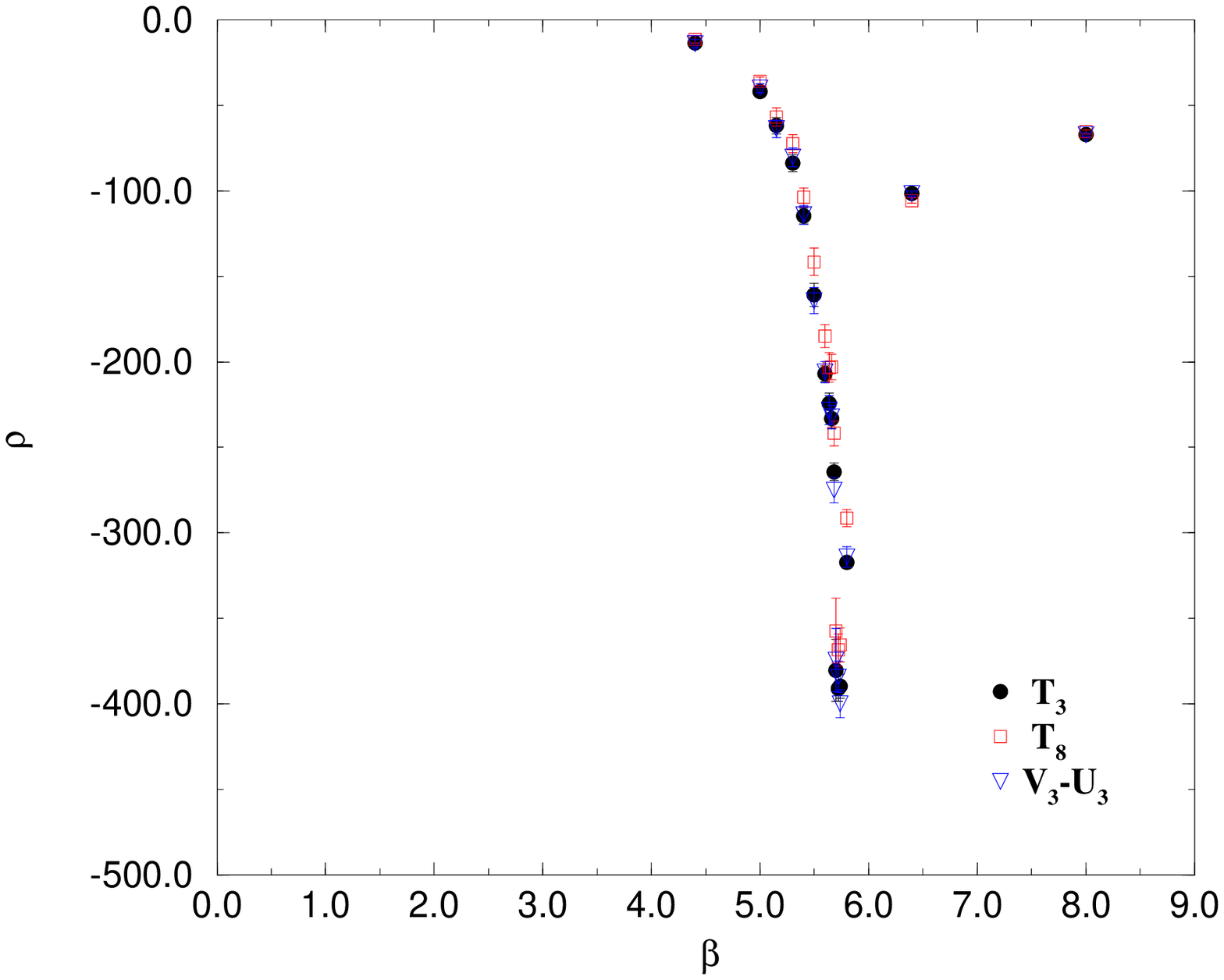}}
\end{minipage}\\
{\bf Fig.2} $\rho$ for the two different monopole species
in the Polyakov projection. $SU(3)$.
\vskip0.1in\par\noindent

\par\noindent
\begin{minipage}{0.95\linewidth}
\epsfxsize0.95\linewidth
{
\epsfbox{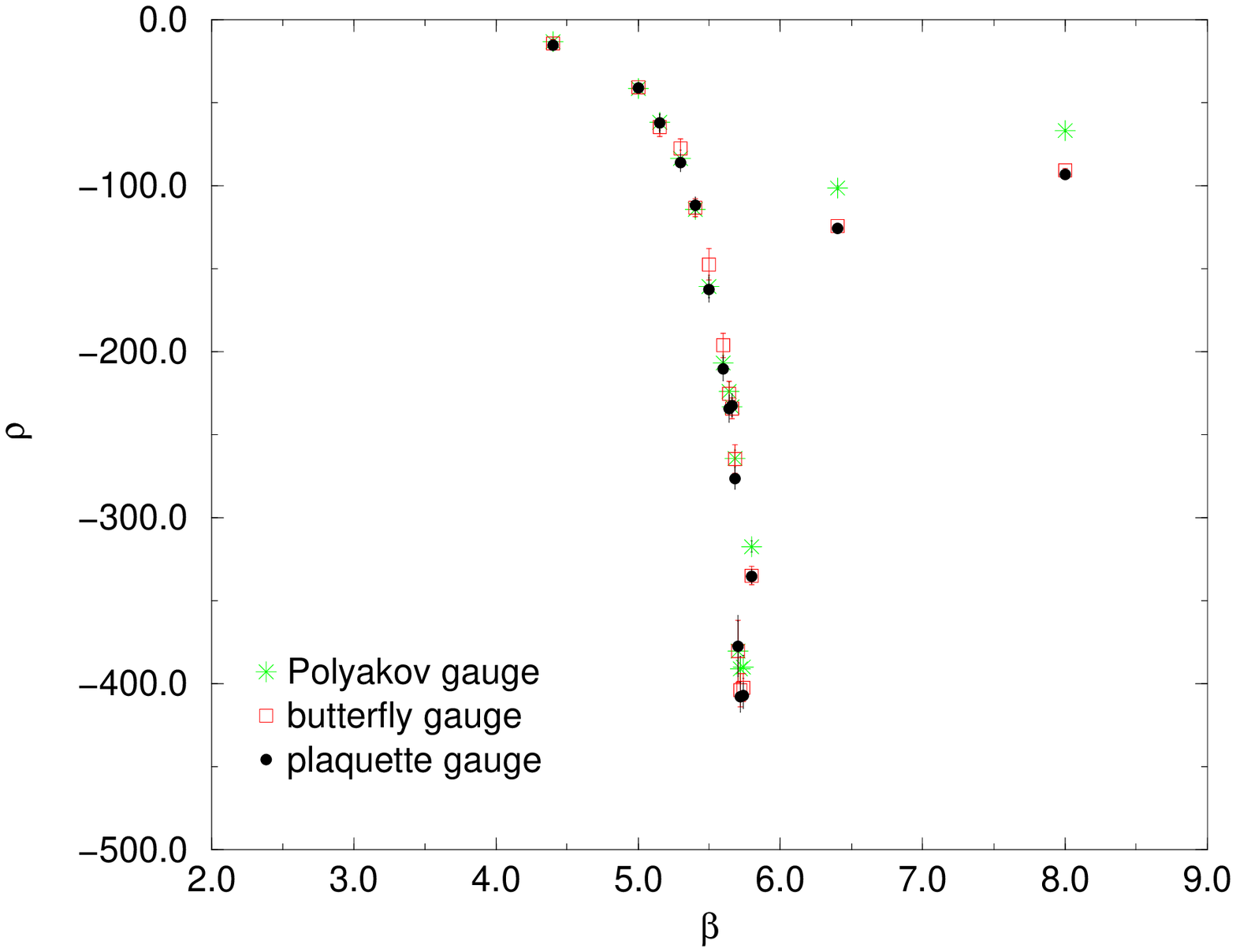}}
\end{minipage}\\
{\bf Fig.3} $\rho$ for different abelian
projections. $SU(3)$.
\vskip0.1in\par\noindent

This implies by eq.(\ref{eq:11})
\[ \mu = \Phi(L^{1/\nu}(\beta_c-\beta))\]
or
\begin{equation}
\frac{\rho}{L^{1/\nu}} = f(L^{1/\nu}(\beta_c-\beta)) \label{eq:13}\end{equation}
The scaling law is obeyed for the appropriate values of $\nu$ and $\beta_c$.
Fig.4 shows how scaling works. The output for $SU(2)$ is $\nu = .62\pm.02$ to
be compared with the expectation, the critical index of 3d Ising model
$\nu = .631(1)$.

For $SU(3)$ we find a similar value, contrary to the expectation which should
be 1/3. However our volumes are not sufficiently large, and further
investigations are on the way.
\par\noindent
\begin{minipage}{0.95\linewidth}
\epsfxsize0.95\linewidth
{
\epsfbox{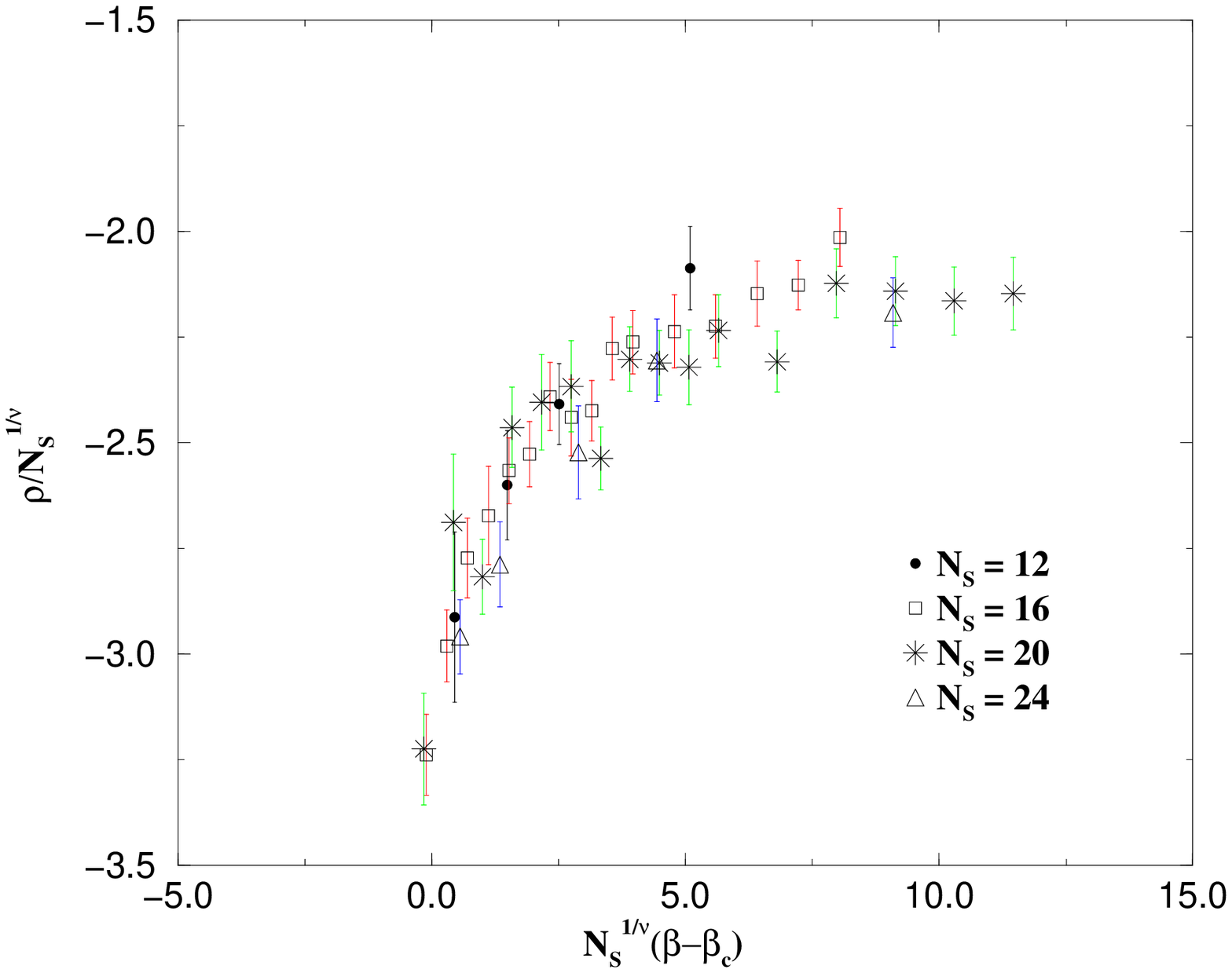}}
\end{minipage}\\
{\bf Fig.4} Finite size scaling eq.(\ref{eq:13}). $SU(2)$.
\vskip0.1in\par\noindent
\section{Conclusions}
A disorder parameter can be defined to investigate condensation of monopoles
in the vacuum of QCD.

QCD vacuum is a dual superconductor. Different monopole species look
equivalent, and condense in connection with confinement, in agreement with
the
conjecture of t'Hooft's\cite{6}.

This is  an important information on the symmetry of vacuum, which
must be explained by any model of confinement.

\end{document}